# Accelerating the Design of Resorbable Magnesium Alloys: A Machine Learning Approach to Property Prediction

Vickey Nandal[1*], Vít Beneš[1,2], Pavel Baláž[1], Jiří Ryjáček[2], Karel Tesař[1,2]

[1]FZU - Institute of Physics, Czech Academy of Sciences, Na Slovance 2, Prague, 182 21, Czech Republic

[2]Department of Materials, Faculty of Nuclear Sciences and Physical Engineering, Czech Technical University in Prague, Trojanova 13, Prague, 120 00, Czech Republic

**Abstract**

Resorbable magnesium (Mg) alloys are promising candidates for temporary medical devices due to their biodegradability and favorable mechanical properties. To accelerate the design of diluted Mg alloys for implants, we developed a data-driven framework to elucidate the complex relationships between composition, processing, and mechanical properties. The framework screens mechanical properties within biocompatible compositional limits, treating degradation as a design constraint rather than an explicit prediction target. Using a dataset of 410 samples, we trained six different machine learning (ML) models to predict yield strength, ultimate tensile strength, and elongation. Among them, ensemble models, particularly CatBoost, demonstrated high predictive accuracy ($R^2$, YS ≈ 0.950, UTS ≈ 0.916 and El ≈ 0.903). SHapley Additive exPlanation analysis revealed that thermomechanical processing conditions and alloying elements such as Zn, Mn and Gd are the most influential factors governing mechanical behavior in diluted Mg alloys. Validation on the experimental dataset confirmed the models' robustness and generalization capability in capturing process-property relationships. The optimized CatBoost model was further employed to generate predictive property maps visualizing the strength-ductility trade-off as a function of Zn-Mn composition. This work establishes a validated ML framework for rapid in silico screening and targeted design of next-generation resorbable Mg alloys.

## 1. Introduction

Magnesium (Mg) and its alloys have gained significant attention as a novel class of biodegradable materials for temporary implants, such as screws, plates, and stents, owing to their favourable biocompatibility and in vivo degradability [1,2]. Their key advantage is their ability to be resorbed by the body over time, eliminating the need for secondary removal surgeries, which reduces patient trauma and healthcare costs. Their elastic modulus and density are closer (41 – 45 GPa, 1.74 – 2.0 g/cm$^3$) to that of natural bone (3 – 20 GPa, 1.8 – 2.1 g/cm$^3$) compared to traditional permanent implants like titanium or stainless steel, which can mitigate the stress shielding effect [3].

However, the widespread clinical adoption of Mg alloys is challenging. The primary obstacle is achieving an optimal balance between mechanical integrity, controlled degradation rate, and biocompatibility [4]. The mechanical properties such as yield strength (YS), ultimate tensile strength (UTS), and elongation (i.e., ductility) must be sufficient to support the healing tissue during its recovery period. These properties are governed by a complex, non-linear interplay between the alloying elements and the thermomechanical processing history (e.g., extrusion, heat treatment) [5]. Traditional trial-and-error experimental approaches to explore this vast design space are time-consuming, expensive, and often yield only incremental improvements.

In recent years, data-driven approaches utilizing machine learning (ML) have been applied in the materials science field to accelerate the materials discovery and design [6–10]. Significant research has demonstrated that ML algorithms, such as Artificial Neural Network (ANN) [11], eXtreme Gradient Boosting (XGBoost) [12] and Monte Carlo tree search (MCTS) [9] for designing the novel alloys and heat treatment processing. By learning from existing experimental data, ML models can capture the important relationships between input variables (composition, processing conditions) and output target properties, enhancing predictive accuracy for novel material compositions. Nevertheless, the intrinsic complexity of ML models frequently constrains their interpretability. Within the paradigm of interpretable machine learning, feature importance analysis employs advanced techniques such as SHapley Additive exPlanation (SHAP) to enable a transparent and quantitative assessment of individual feature contributions towards the predictions [13]. This paradigm can significantly reduce the number of required experiments by guiding researchers toward the most promising regions of the design space.

While several studies have applied ML to predict the mechanical properties of Mg alloys [14–16], this work establishes a more comprehensive framework in the field of bioresorbable Mg alloys where biocompatibility is a limiting criterion for the final alloy composition. In contrast to previous ML studies on Mg alloys [14–16], which focused primarily on structural alloy optimization, the present work explicitly focuses within a biomedical compositional design window constrained by cytotoxicity considerations and provides interpretable SHAP-guided property maps for alloy screening. Unlike conventional structural alloys, the compositional design space of bioresorbable alloys is highly restricted due to the potential cytotoxicity of alloying elements [4]. This limitation applies not only to non-essential elements but also to essential biogenic elements when present above physiologically acceptable ranges. For instance, manganese (Mn) is beneficial at trace concentrations but may become toxic when their accumulation exceeds tolerance limits, while magnesium remains physiologically safe throughout the concentration range. Importantly, this toxicity threshold is strongly influenced by the in vivo degradation rate of the implant, as rapid corrosion may lead to locally elevated ion release that surpasses biocompatibility limits. Consequently, the integration of ML with biocompatibility-informed compositional constraints, combined with interpretable SHAP-based visualization tools, is essential for guiding the safe and effective development of Mg-based bioresorbable alloys.

In this work, we have compiled a diverse dataset that includes 14 compositional elements promising for implant applications in Mg alloys (for example, Zn, Mn, Y, Gd etc.) and 4 critical, yet often overlooked, thermomechanical processing parameters such as preheat temperature (i.e., homogenization temperature), extrusion temperature and extrusion ratio. We conducted a rigorous evaluation of six distinct ML models, from simple linear regression to advanced ensemble methods, to systematically identify the most suitable algorithm for reasonable predictions for mechanical properties. The degradation behavior was not explicitly modeled due to the limited availability of standardized and comparable corrosion datasets, as degradation rates are highly sensitive to testing environment, surface condition, and in vivo variability. After selecting the best prediction model for property predictions, we validated it using an unseen experimental dataset that was not used during training and testing. At the end, the optimized model was also used to generate predictive maps that directly address the challenge of maximizing strength and ductility within selected composition interval. This systematic approach provides a robust and validated technique/tool for accelerating the design of next-generation resorbable diluted Mg alloys.

In this study, we develop a comprehensive ML framework to predict the mechanical properties of resorbable Mg alloys, as illustrated in Fig. 1. The ML workflow includes several steps. We collect the large dataset of resorbable Mg alloys from the peer-reviewed articles in literature, perform extensive data analysis to understand its characteristics, and train six different ML models to predict YS, UTS, and elongation. We evaluate their performance, identify the most influential features, and use the best-performing model to create predictive maps for guiding the design of new alloys with targeted properties. In addition, we also validated the predictive maps using the experimental dataset.

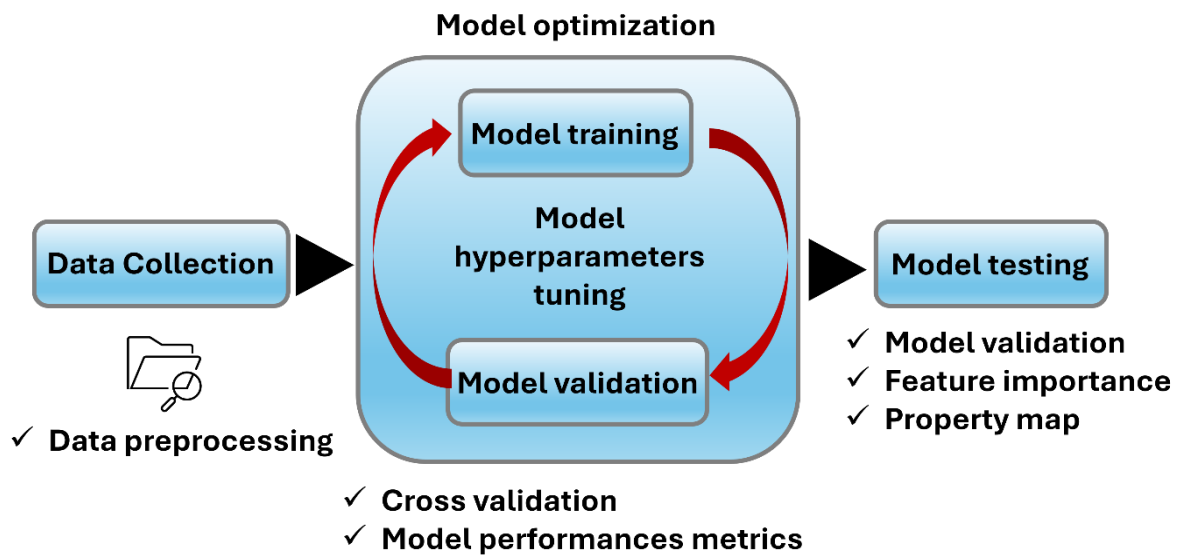

**Fig. 1:** Description of a comprehensive ML framework which includes data collection, model validation, machine learning model evaluation and designing of diluted Mg alloys.

2. Methods
2.1. Data collection and data analysis

The raw dataset for this study was compiled from the peer-reviewed articles for various resorbable Mg alloy systems at different processing conditions, which consisted of 312 experimental data samples. The dataset references can be found in the Zenodo repository. To ensure data quality and consistency, inclusion criteria were established in which data were limited to extruded or preheat treated conditions (i.e., homogenised) combined with extruded conditions in Mg alloys, and mechanical properties had to be obtained from room-temperature uniaxial tensile tests. To ensure uniformity across samples, we represented multistep preheat treatment using the maximum reported temperature and the total duration of the heat treatment process. The raw dataset was augmented from 312 samples to 600 samples using the mean and standard deviation of the measured mechanical properties. In addition, data cleaning was performed to resolve inconsistencies, leading to the removal of 190 data points (i.e., removed from the collected dataset size of 600 samples) with incomplete processing or target information. The final dataset contains 410 unique data points, each comprising 21 attributes such as 14 compositional features (wt.%), 4 processing parameters, and 3 target mechanical properties (YS, UTS, and El). The characteristics of the dataset are fundamental to the development of the ML models. The ranges of attributes are shown in Table 1, confirm that our dataset ranges a wide and diverse materials space, which is essential for training generalizable models.

**Table 1.** List of chemical composition, attributes and outputs (targets) of magnesium alloys.

| No | Description (Unit) [Abbreviation] | Min | Max | Range |
|---|---|---|---|---|
| 1 | Mg (wt. %) | 90.38 | 100 | 9.62 |
| 2 | Mn (wt. %) | 0.00 | 2.00 | 2.00 |
| 3 | Sr (wt. %) | 0.00 | 1.60 | 1.60 |
| 4 | Dy (wt. %) | 0.00 | 2.00 | 2.00 |
| 5 | Al (wt. %) | 0.00 | 3.10 | 3.10 |
| 6 | Ca (wt. %) | 0.00 | 5.00 | 5.00 |
| 7 | Zn (wt. %) | 0.00 | 7.60 | 7.60 |
| 8 | Gd (wt. %) | 0.00 | 9.62 | 9.62 |
| 9 | Nd (wt. %) | 0.00 | 3.00 | 3.00 |
| 10 | Y (wt. %) | 0.00 | 7.25 | 7.25 |
| 11 | Sn (wt. %) | 0.00 | 5.86 | 5.86 |
| 12 | Ce (wt. %) | 0.00 | 0.39 | 0.39 |
| 13 | La (wt. %) | 0.00 | 0.32 | 0.32 |
| 14 | Zr (wt. %) | 0.00 | 1.00 | 1.00 |
| 15 | Extrusion temperature (K) [Extr. Temp.] | 298 | 743 | 445 |
| 16 | Extrusion speed (m/min) [Extr. Speed] | 0 | 50 | 50 |
| 17 | Extrusion ratio [Extr. Ratio] | 1 | 81 | 80 |
| 18 | Preheat temperature (K) [Preheat Temp.] | 298 | 813 | 515 |

| No | Description (Unit) [Abbreviation] | Min | Max | Range |
|---|---|---|---|---|
| 19 | Preheat time (h) [Preheat Time] | 0 | 106 | 106 |
| 20 | Yield strength (MPa) [YS] | 33 | 387 | 354 |
| 21 | Ultimate tensile strength (MPa) [UTS] | 73.00 | 412.80 | 339.80 |
| 22 | Elongation (%) [El] | 0.50 | 51.00 | 50.50 |

Detailed distribution of all attributes, including alloying elements and processing parameters, is presented as histograms in Supplementary Fig. S3. Many alloying elements, such as Mn, Sr, and Al, have right-skewed distributions, indicating their presence as minor additions in most alloys. The mechanical properties (YS, UTS, El) exhibit broader, more centrally distributed patterns, reflecting the wide range of performance outcomes in the dataset.

In order to ensure accurate prediction, it is important to determine the optimal set of input features separately for each mechanical property considered in this work. As the properties such as ultimate tensile strength, yield strength, and elongation are governed by different underlying factors, a property-specific feature selection strategy is required. For each dataset corresponding to a given mechanical property, the Pearson Correlation Coefficient (PCC) was first applied to quantify the linear dependence among features. A correlation analysis was conducted to assess the relationship between pairs of all the variables using the PCC heat map, as indicated in Fig 2. The correlation coefficient ranges from -1 to +1, where -1 and +1 indicate perfect negative and positive correlations, respectively, and 0 denotes no linear relationship. The correlations between individual input features and the target properties vary from generally weak to strong. For example, UTS and YS have a positive correlation with Zn as well as with Mn. Elongation (El) exhibits a positive correlation with Gd, making Gd the alloying element most positively correlated with elongation among all considered elements. The Gd and Y has negative correlation with YS and UTS, as shown in Fig. 2.

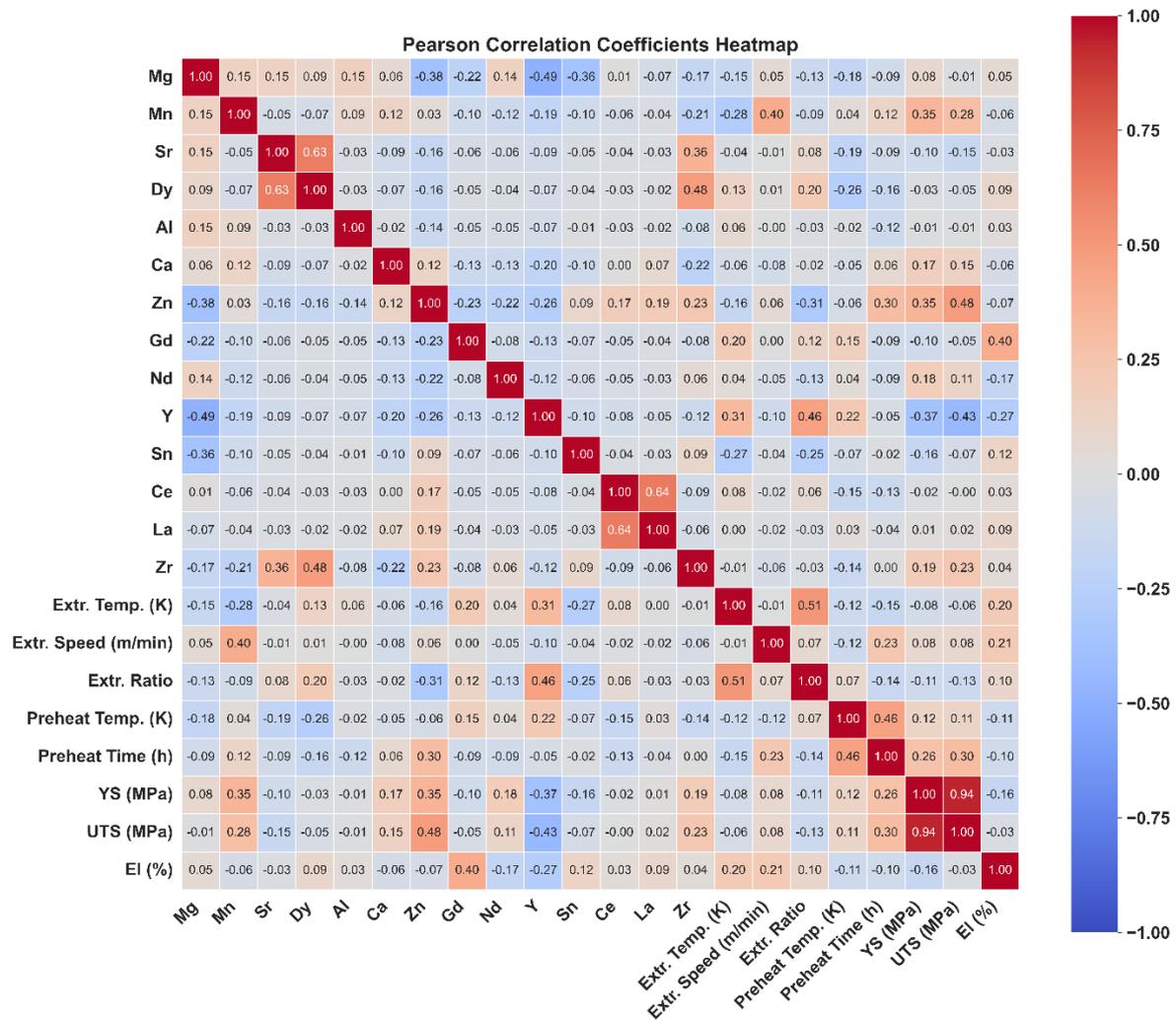

**Fig. 2:** Pearson correlation coefficients between the attributes (alloys elements and processing parameters) and target properties. Higher color intensity corresponds to stronger correlations between the respective features.

### 2.2. Machine learning models selection and optimization

In this study, a comparative analysis was conducted on six different machine learning models to evaluate their predictive performance on all the target properties. All machine learning models were implemented using Python 3.11.0 in a Jupyter Notebook environment. The models were developed using the Scikit-Learn library, along with the XGBoost and CatBoost packages for gradient boosting. Matplotlib and Seaborn were used for visualization. The selected ML algorithms ranged in complexity from traditional to advanced ensemble methods. For instance, the foundational models included linear regression, a baseline linear approach, K-Nearest Neighbors (KNN), a non-parametric instance-based method, and Decision Tree (DT), a rule-based non-linear model. In addition, three ensemble models were evaluated, Random Forest (RF) [17], an ensemble of decision trees based on bagging and two state-of-the-art gradient boosting frameworks, XGBoost [18] and CatBoost [19]. The advantages, limitations, and suitability of all models are given in Supplementary Section 1.

For ML analysis, the complete dataset was partitioned into a training set (80%) and a testing set (20%) for all the models. The testing set, i.e., dataset which does not take part in the model training, was held out and used only for the final evaluation of the trained models to ensure an unbiased assessment of their generalization performance. Given the varying scales and units of the input features (e.g., wt.% for composition, K for temperature), all features were standardized using the StandardScaler, which transforms the data to have a mean of zero and a standard deviation of one.

In addition, model hyperparameters were tuned using a Grid Search with 5-fold cross-validation to optimize predictive capability and prevent overfitting, as illustrated in Supplementary Fig. S4. Cross-validation generally reduces the bias in performance estimation by averaging results over multiple splits. This method exhaustively searches a parameter space to identify the combination that yields the highest performance. The final, optimized hyperparameters for each model and for each of the three target mechanical properties are listed in Supplementary Table S1.

**2.3. Performance evaluation of learning models**

To quantitatively assess the performance of all models, we employed three standard regression metrics such as coefficient of determination ($R^2$), mean absolute error (MAE), root mean squared error (RMSE), which are given as the following equations (1), (2) and (3), respectively:

$$R^2 = 1 - \frac{\sum_{i=1}^{n}(y_i - \hat{y}_i)^2}{\sum_{i=1}^{n}(y_i - \bar{y}_i)^2} \tag{1}$$

$$MAE = \frac{1}{n}\sum_{i=1}^{n}|y_i - \hat{y}_i| \tag{2}$$

$$RMSE = \sqrt{\frac{1}{n}\sum_{i=1}^{n}(y_i - \hat{y}_i)^2} \tag{3}$$

where $\hat{y}_i$ is the predicted value for the *i-th* sample, $y_i$ is the corresponding actual value, $\bar{y}$ is the mean of all actual values, and *n* is the total number of samples. The $R^2$ indicates the proportion of variance explained by the model, MAE reflects the average magnitude of prediction errors and RMSE penalizes larger errors by taking the square root of the mean squared differences between predicted and actual values, as given in corresponding equations.

**2.4. Model validation and predictive mapping**

To evaluate the generalizability of the trained models beyond the curated training dataset, we performed a thorough validation. For example, an independent validation set was created by extracting the relevant experimental data from literature. This dataset was entirely excluded from model training, tuning, and initial testing phases. The optimized machine learning model was then applied to the validation dataset to observe the accuracy of model.

In addition, to visualize the model's predictions and guide the alloy design, the optimized CatBoost model was employed to generate predictive maps. These were shown as 2D contour plots, illustrating the predicted mechanical properties as a function of key alloying elements.

During the generation of these plots for two major elements, all other input features, including the remaining chemical compositions were kept constant at their respective mean values.

## 3. Results and Discussion
### 3.1. Performance of machine learning models

The performance of the prediction results which correspond to the six different ML algorithms such as linear, KNN, XGBoost, CatBoost, Decision tree and Random Forest are summarized in Fig. 3. Most of the ML models except simpler one (i.e., linear model) exhibited excellent predictive performance ($R^2 > 0.88$) for YS and UTS, as shown in Fig. 3a and 3b, respectively. In the case of elongation, it is clearly indicating that the ensemble models (Random Forest, CatBoost, XGBoost) significantly outperform the other considered models, as indicated in Fig. 3c.

For instance, to predict the YS (Fig. 3a), the CatBoost model achieved the highest $R^2$ of 0.950, followed closely by XGBoost ($R^2$=0.944) and Random Forest ($R^2$=0.936). For UTS prediction (Fig. 3b), the performance was similar, with CatBoost ($R^2$=0.916), XGBoost ($R^2$=0.911), and Random Forest ($R^2$=0.905) showing the best prediction results. For predicting elongation (Fig. 3c), which is often more difficult to model due to its high sensitivity to microstructure and defects, the ensemble model again proved superior, with CatBoost model achieving the highest $R^2$ values of 0.903. Overall, the CatBoost model demonstrated a superior performance of all three target properties, with the lowest MAE and RMSE values in most cases. Hence, it was selected as the final model (i.e., optimized model) for further analysis and predictive mapping application in this work.

The high predictive accuracy of the CatBoost model is further visualized in the individual target property plots, as shown in Fig. 4. These plots compare the model-predicted values against the actual experimental values for YS, UTS, and El for both the training and testing datasets. The data points are tightly clustered around the ideal 45º line (y = x), indicating strong agreement. The high $R^2$ values for both the training set (e.g., 0.96 for YS) and the testing set (e.g., 0.95 for YS) confirm that the model is not only accurate but also generalizes well to new data without significant overfitting.

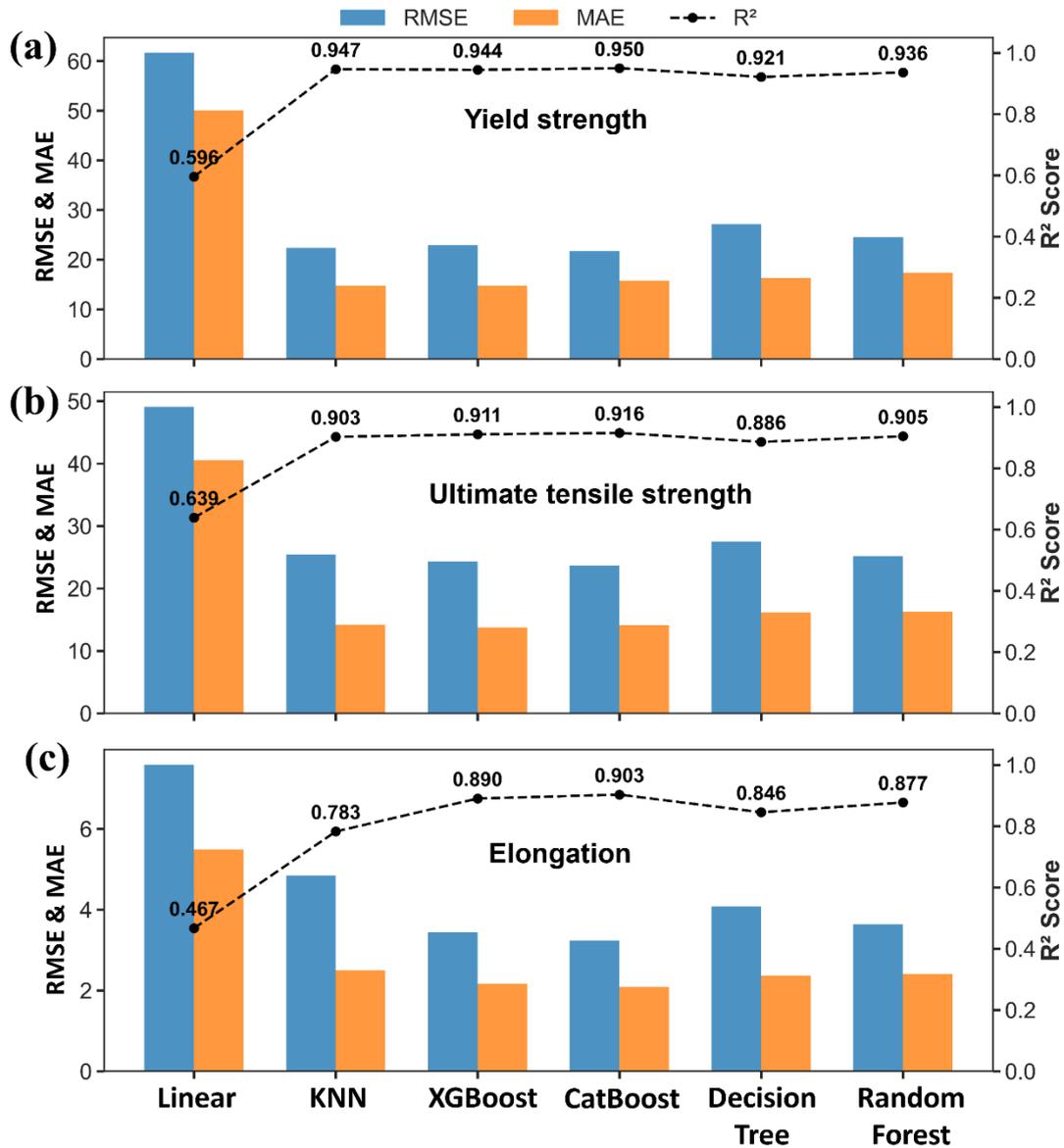

**Fig. 3:** Comparison of performance evaluation including the $R^2$, MAE and RMSE, for all the considered models on the testing dataset.

It should be noted that despite generally strong performance metrics ($R^2$, RMSE, MAE), close examination of the predicted versus experimental values, particularly for YS, UTS and elongation, reveals instances where the model's predictions align along a near-horizontal line (around predicted UTS of 175 – 185 MPa) for a range of experimental values, as indicated in Figure 4b. This near-horizontal prediction region indicates a tendency of the model to regress toward an average response within this property range, reducing sensitivity to variations in experimental mechanical strength values. This behavior may arise from limited data density in specific regions of the feature space or from insufficient representation of strengthening mechanisms in the input descriptors. For instance, this pattern suggests specific limitations in the model's ability to fully capture the underlying mechanisms and microstructural features such as average grain size, crystallographic texture, dislocation density, and distribution of precipitates, governing these properties in certain regimes.

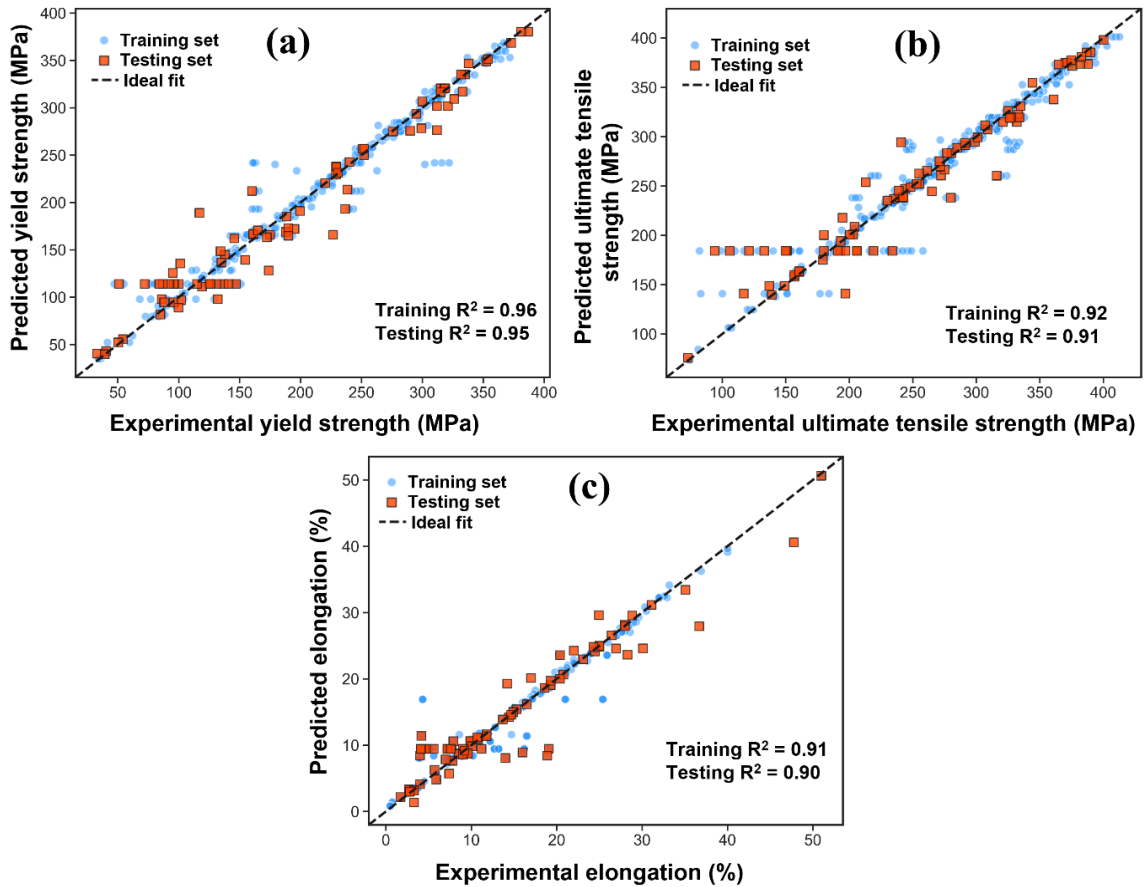

**Fig. 4.** Predicted and experimental (a) YS, (b) UTS and (c) El for the CatBoost model (i.e., optimized model) in the training and testing datasets.

### 3.2. SHAP-based feature analysis

In present study, to further understand the influence of individual features (including the process parameters and alloying elements) and model's predictions on mechanical properties (i.e., yield strength, ultimate tensile strength, and elongation), a SHapley Additive exPlanation (SHAP) methodology is utilized using the optimized CatBoost model. It enables a quantitative assessment of each feature's contribution to the mechanical properties of Mg alloys. As shown in Fig. 5 (a, c, e), the horizontal axis represents the SHAP values, the vertical axis corresponds to the features, and each point represents a sample. A positive or negative SHAP value for a given feature indicates that the feature respectively enhances or weakens the mechanical properties. The magnitude of each feature value is represented by the color of the corresponding point. The importance of a feature is reflected by the width of its horizontal spread, where a wider distribution indicates a greater impact on the prediction results. As shown in Fig. 5 (b, d, f), the mean absolute SHAP value of each feature represents its relative importance.

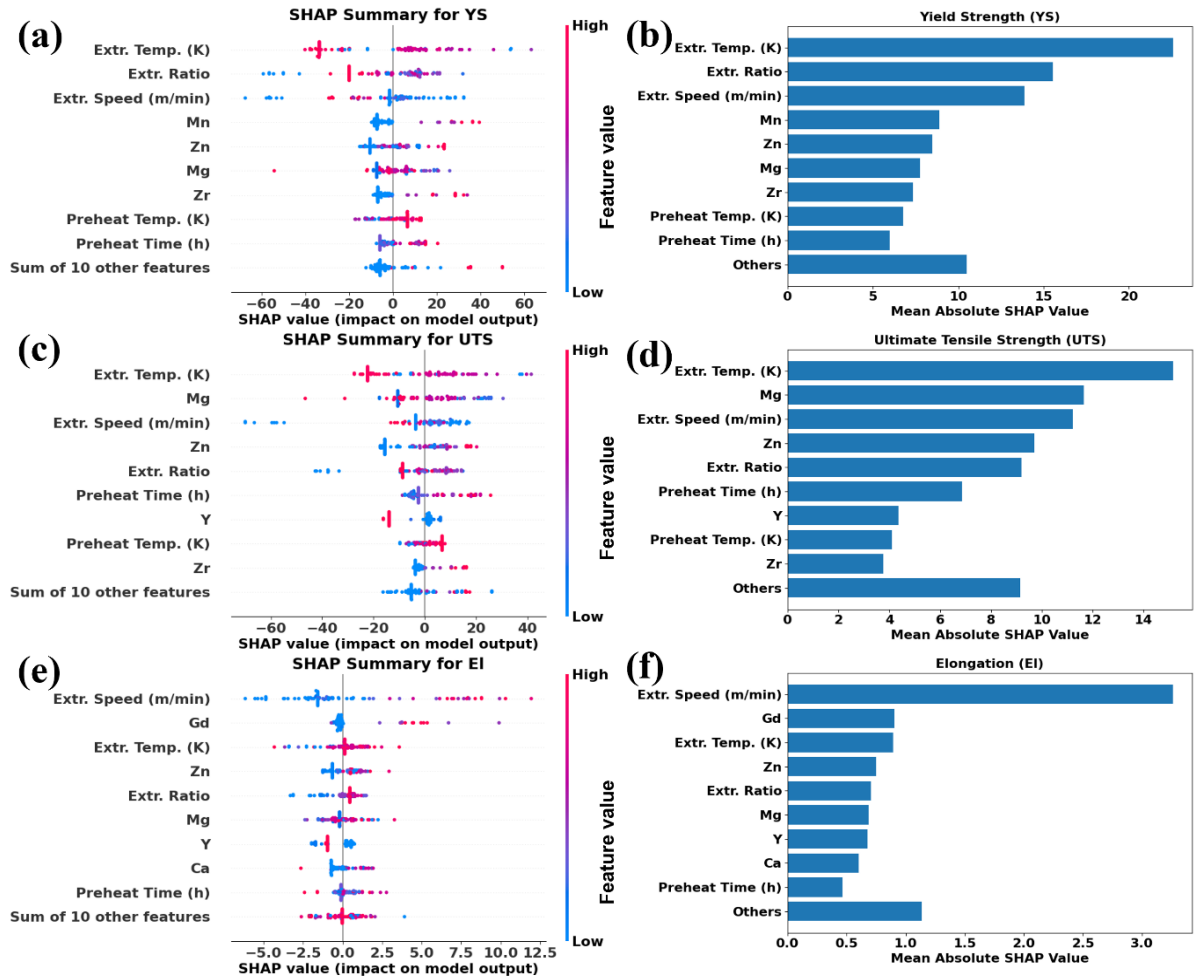

**Fig. 5.** Evaluation of importance of feature on mechanical properties in Mg alloys: Distribution of SHAP values for (a) yield strength, (c) ultimate tensile strength, (e) elongation; Mean absolute SHAP values for (b) yield strength, (d) ultimate tensile strength, (f) elongation.

For both YS and UTS, a clear trend was observed where thermomechanical processing parameters and specific alloying elements in Mg matrix such as Zn and Mn are the most dominant features, as shown by the spread and coloring of SHAP values in Fig. 5a and 5c and their high mean absolute SHAP values in Fig. 5b and 5d, respectively. For instance, the local SHAP plot, which can be viewed as a horizontal projection of the waterfall plot, groups the features that increase the prediction (red) and decrease the prediction (blue). The plot for an alloy is shown in Supplementary Fig. S5, indicates that Zn and Mn contribute most strongly to the positive predictions to the predicted YS and UTS. The addition of these elements in Mg alloys increases the mechanical properties via grain size reduction, precipitation hardening and solid solution strengthening [20]. For instance, the SHAP results are consistent with typical strengthening mechanisms in Mg alloys. The extrusion ratio provides the mechanical strain necessary to trigger dynamic recrystallization (DRX), creating the nucleation sites required to achieve finer grain structure. Simultaneously, extrusion temperature provides the thermal activation energy required to facilitate the kinetics of nucleation of grains and boundary migration. The synergy of a high extrusion ratio and an optimized processing temperature results in significant grain refinement, consistent with the Hall-Petch relationship [21]. Among the alloying elements, the significant influence of Y, Zn and Mn on YS and UTS is consistent with their known roles as potent solid solution and precipitation strengtheners in Mg alloys [4]. The positive contribution of Zn is attributable to solid solution strengthening, whereas Zn atoms create lattice distortions that impede dislocation motion [22]. Conversely, Mn contributes primarily through grain refinement mechanisms, specifically particle-stimulated nucleation and Zener pinning, which restrict grain growth during thermomechanical processing [23]. These interpretations validate that the ML model captures physically meaningful process-structure-property relationships rather than purely statistical correlations.

For elongation, Gd and Zn were the most dominant feature among the alloying elements, while extrusion speed and temperature were the most important features based on the SHAP summary in Fig. 5e and the corresponding bar chart (i.e., mean absolute SHAP values) in Fig. 5f. This is because these parameters critically affect the final crystallographic texture and grain structure, which govern the deformation mechanisms and thus the ductility of hexagonal close-packed metals like magnesium alloys [4]. For instance, alloying with Al plays a key role in strengthening commercial Mg alloys (e.g., AZ91, AZ31) through solid solution strengthening and precipitation hardening via ($Mg_{17}Al_{12}$) [24]. However, its use in biomedical applications is limited, as Al has been associated with neurotoxicity and a possible link to Alzheimer's disease, raising significant biocompatibility concerns [25].

### 3.3. Validation of optimized predictive ML model

In the section 3.1, the test dataset was randomly sampled from the complete dataset and was not involved in the training of the machine learning models. To demonstrate the generalizability of our ML framework, the best-performing model (CatBoost) was further validated with an entirely new, independent dataset. This external validation set was curated by extracting experimental data for different alloy systems from several peer-reviewed literature articles. Crucially, this dataset comprised both Mg-Zn-Mn [26–28] ternary alloys with diverse compositions and thermomechanical processing parameters, as listed in Table 2. This approach was designed to challenge the models' predictive capabilities across different chemical composition spaces and confirm that they were not overfitted to the initial training data, thereby demonstrating their robustness for practical alloy design for bioresorbable applications.

The CatBoost model's performance on the unseen Mg-Zn-Mn alloy system, i.e., dataset which does not take part in the model training and testing, as illustrated in Fig. 6, suggests promising generalization through preliminary validation. In the case of yield strength (Fig. 6a), the model's predictions for YS were also highly accurate. For Instance, the predicted YS for the Mg-1Zn-1Mn alloy (242.55 MPa) and Mg-6Zn-0.3Mn alloy (150.34 MPa) were nearly similar to their experimental counterparts (246 MPa and 152 MPa, respectively). Across all four alloys in this system, the YS predictions closely agreed with the experimental results, confirming the model's ability to generalize to new compositions within the Mg-Zn-Mn alloy system. The property map values (discussed in Section 3.4) for all the three properties also show good agreement with both the experimental and predicted results, as shown in Fig. 6.

**Table 2:** Composition and processing conditions of new experimental data of Mg-Zn-Mn system.

| Alloy composition wt.% | Preheat Temp. (K) | Preheat Time (h) | Extrusion Temp. (K) | Extrusion Speed (m/min) | Extrusion Ratio |
|---|---|---|---|---|---|
| **Mg-2Zn-1Mn** [26] | 553 | 1.5 | 553 | 2.5 | 25 |
| **Mg-2Zn-2Mn** [26] | 553 | 1.5 | 553 | 2.5 | 25 |
| **Mg-6Zn-0.3Mn** [27] | 673 | 1 | 633 | 1.0 | 4.5 |
| **Mg-1Zn-1Mn** [28] | 473 | 1 | 573 | 1.32 | 10 |

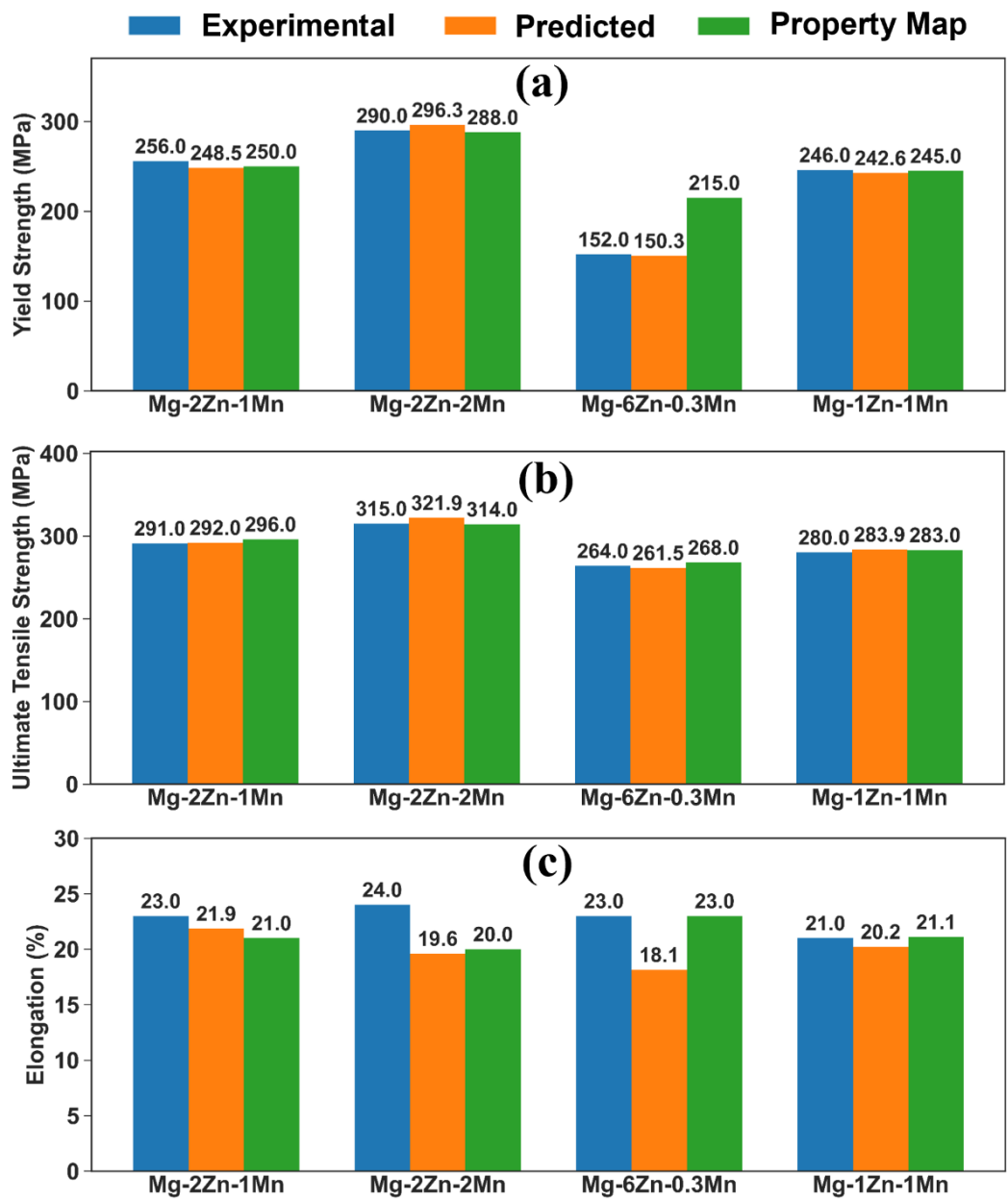

**Fig. 6:** Comparison of optimized CatBoost model's direct prediction, experimental values and property values derived from predictive map for (a) YS (b) UTS and (c) elongation of the new validation dataset from the literature [26–28] for Mg-Zn-Mn alloys.

In the case of ultimate tensile strength (Fig. 6b), the model demonstrated reasonable accuracy in predicting UTS compared to experimental values. For instance, the prediction for the Mg-2Zn-1Mn alloy was 292 MPa, in almost perfect agreement with the experimental value of 291 MPa. Similarly, for the Mg-1Zn-1Mn alloy, the predicted value of 283.88 MPa closely matched the experimental 280 MPa. In most cases, the best-performing model successfully captured the mechanical properties of these alloys across the entire set.

In the case of predicting the elongation as shown in Fig. 6c, it is typically difficult due to its high sensitivity to microstructural features, such as grain size, texture, defect concentration (such as MgO particles), etc. Despite this, the CatBoost model performed reasonably well. The prediction for the Mg-1Zn-1Mn alloy (20.22%) was in excellent agreement with the experimental value (21%). For Mg-2Zn-1Mn, the prediction elongation is very close to the experimental value (Fig. 6c). The model consistently captured the general ductility trends, indicating a robust understanding of the factors influencing elongation.

Application of the optimized ML model (CatBoost) showed strong predictive performance when evaluated on the literature-derived dataset (Fig. 6). Although there are some deviations in predicted and experimental values, the model still has reasonable quantitative accuracy, even when the extremely important microstructural parameters (e.g. grain size, texture) were not included in the model. In the current study, grain size was not included due to its relative difficulty in being determined when compared with other investigated parameters. This successful validation confirms that the models have learned the fundamental process-property relationships and are not overfitted to our specific experimental conditions, highlighting their robustness and potential for broader use. This performance provides the necessary confidence to employ the models for their primary purpose, guiding the efficient design of new alloys with targeted properties.

### 3.4. Predictive mapping for alloy designing

An important application of the trained model is to guide future experiments. Using the optimized CatBoost model, we generated predictive property maps as a function of Zn-Mn contents, promising sets of alloying elements in resorbable Mg alloy systems, as illustrated in Fig. 7 through a series of contour plots. For instance, the Figs. 7(a–c) show the combined influence of Zn and Mn in Mg alloy system. Each plot represents a specific mechanical property, such as yield strength, ultimate tensile strength, or elongation with color gradients indicating the magnitude of the specific property. For instance, yellow color corresponds to higher values, whereas dark blue or purple color indicate the lower values. Typically, the desirable range of mechanical properties for biodegradable implant applications requires a yield strength greater than 200 MPa and an elongation exceeding 15% [29].

In the Mg-Zn-Mn system, Fig. 7a displays YS as a function of Zn (0–7.5 wt.%) and Mn (0–2 wt.%). The YS values range from approximately 206 MPa to 286 MPa, showing a strong positive correlation with Mn content. Notably, YS increases significantly when Mn content exceeds 1.5 wt.%, reaching peak values in alloys with low Zn (<2 wt.%) and high Mn (>1.75 wt.%). Fig. 7b presents the corresponding UTS values, ranging from 259.5 MPa to 313.5 MPa. As with YS, UTS improves with increasing Mn, especially in the high Mn and low-to-intermediate Zn region (0–4 wt.% Zn).

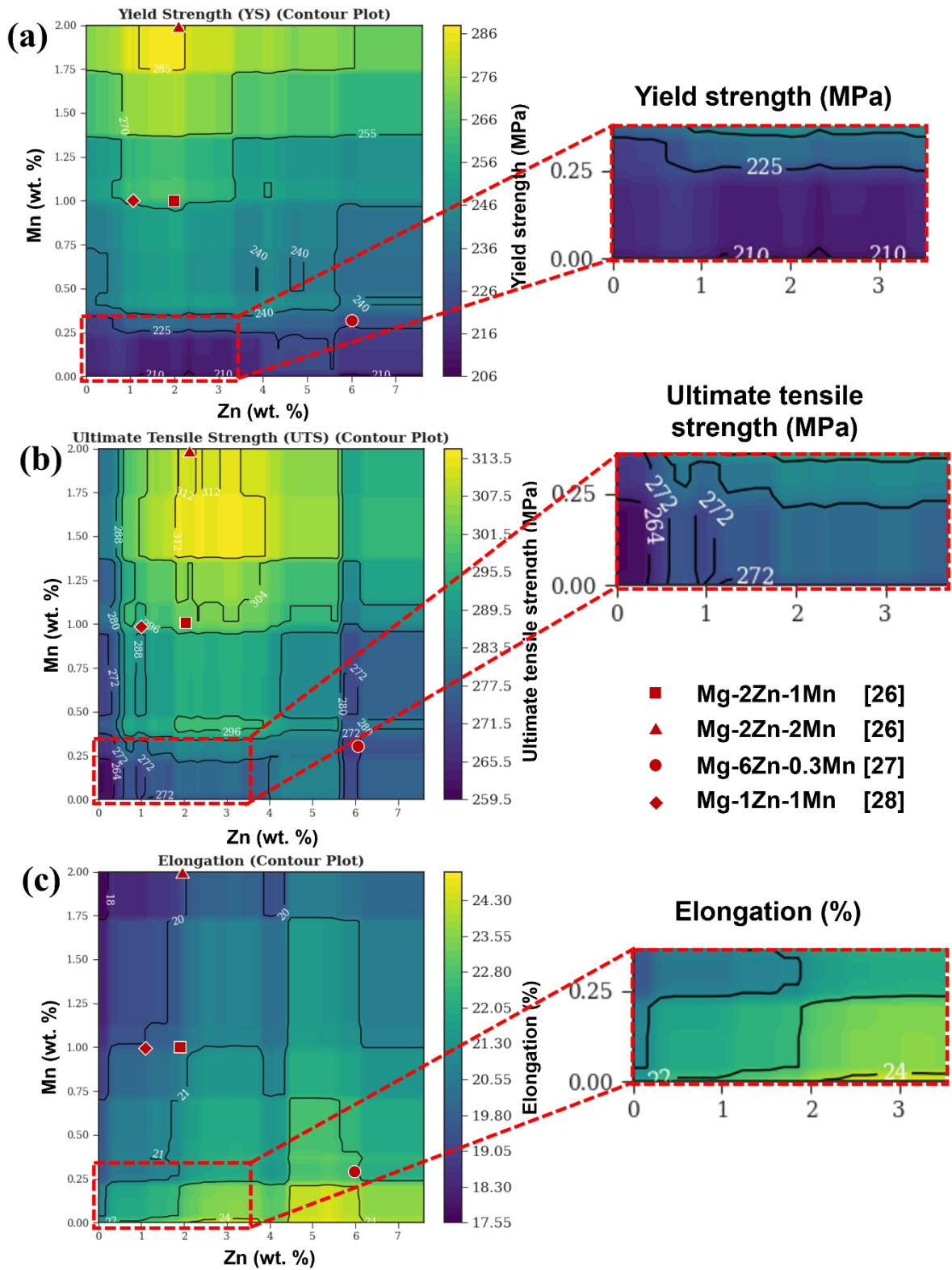

**Fig. 7:** Property prediction maps of varying (a-c) Zn-Mn concentrations and their influence on YS, UTS, and El. Insets indicate the optimized range for low-alloyed Mg-Zn-Mn alloys with acceptable strengthening range for resorbable applications.

Fig. 7c highlights the elongation (i.e., ductility), which varies between 17.55% and 24.30%. Here, an inverse relationship with strength is evident, higher strength (associated with high Mn) corresponds to lower elongation (<19%), while the most ductile alloys are found at low Mn (<0.5 wt.%) and moderate Zn concentrations (0.5–6 wt.%). This trend demonstrated the typical strength-ductility trade-off inherent to the Mg-Zn-Mn alloy system. It is interesting to note that the experimental samples presented in Section 3.3 are also plotted on the predictive maps (Fig. 7), showing a close agreement between the experimentally measured and model-predicted results. For better comparison and visualization, the property map predictions are compared with the best performed ML model (i.e., CatBoost) and experimental samples in Fig. 6, indicating the reasonable agreement.

The discrepancy between the direct CatBoost predictions (Fig. 6) and the property map values (Fig. 7) results from differences in the prediction methodologies employed. The property maps are generated by systematically varying the selected compositional variables while fixing all remaining input features, including thermomechanical processing parameters, to their respective dataset mean values (discussed later in this Section). Consequently, the maps represent averaged compositional trends under standardized processing conditions. However, this simplification may not fully capture the correlations that exist among the various input variables. This might miss some correlations between different variables. In contrast, the direct predictions shown in Fig. 6 are obtained using the actual experimental processing parameters (refer to Table 2) of each alloy, thereby capturing processing condition-specific predictions. For instance, in Fig. 6(a), a difference of 64.7 MPa for YS is observed between the map-derived value (215 MPa) and the model's direct prediction (150.3 MPa). These deviations underscore the strong coupling between composition and processing in determining mechanical performance and confirm that the property maps should be interpreted as generalized compositional guidance rather than exact condition-dependent predictions.

In addition, the primary challenge of these alloys is their potential use for biomedical implants. Poor corrosion resistance, often connected with the presence of intermetallic phases, can lead to the premature loss of mechanical integrity before the tissue has healed fully. Furthermore, while alloying can optimize the mechanical strength and degradation rate, the selection of alloying elements and their concentration must be carefully considered to avoid cytotoxicity and possible hypersensitivity reactions [20,30]. For example, if a target application requires a UTS of ~260 MPa, YS of ~200 MPa with an elongation of at least 20% a weighting for the lowest acceptable concentrations of alloying elements, with respect to their individual toxicity, should be taken into consideration. For instance, the red dotted frames in Fig. 7 are the area recommended by the contour maps for a detailed study of diluted Mg alloys (i.e., low alloyed and excellent mechanical properties suitable for biomedical implants). The corresponding insets show the magnified views of these highlighted regions, emphasizing the diluted Mg alloy domain.

It should be noted that the degradation behavior directly governs both mechanical stability and biological (i.e., in vivo) response. Excessively rapid corrosion may result in premature failure of implant and localized accumulation of degradation products (e.g., $Mg^{2+}$ ions, hydrogen gas), potentially exceeding cytocompatibility thresholds and causing harmful reactions. Conversely, overly slow degradation may prolong the presence of the implant beyond the intended designed window for temporary implant. Therefore, optimal implant design requires a balance between mechanical integrity and degradation kinetics within physiologically acceptable ion-release limits. In this work, mechanical property predictions and the property maps are conducted within biocompatibility-informed compositional space for bioresorbable Mg alloys, while degradation rate predictions are planned as a follow-up work.

A key assumption in generating these maps is that all input features are not being varied while Zn-Mn elements are plotted against the mechanical properties (i.e., all elements other than Zn, Mn and Mg, and all processing parameters were held constant at their dataset mean values). For features with skewed or non-unimodal distributions, as seen in Supplementary Fig. S3, the mean may not represent a physically realistic or common condition. Consequently, the absolute values on the predictive maps should be interpreted as important indicators of property trends within the model space, rather than exact quantitative predictions for a specific alloy. This approach serves to visualize the model's learned relationships, providing directional guidance for experimental design. Future work could expand on using these ML models and predictive maps to produce novel resorbable alloys and expand property maps to a more targeted property prediction.

## 4. Conclusion

We have developed and validated a machine learning framework to accelerate the design of diluted magnesium alloys, primarily considered for resorbable implants. Our work shows that ensemble models, particularly CatBoost, can accurately predict yield strength, ultimate tensile strength and elongation ($R^2 > 0.90$) by learning the complex relationships between alloy composition, processing parameters, and mechanical performance. The framework's strong predictive capability was confirmed against an entirely unseen set of literature data, proving it is well-generalized and not overfitted. The optimized model has achieved acceptable accuracy. The resulting models can serve as a significant tool for the inverse design of alloys, providing a data-driven methodology to guide the optimization and possibly breaking of the strength-ductility trade-off. This approach can significantly narrow the experimental search space, accelerating the development of next-generation biodegradable implants with tailored mechanical properties.

**Declaration of Competing Interest**

The authors declare that they have no known competing financial interests or personal relationships that could have appeared to influence the work reported in this paper.

**Acknowledgements**

This work, led by V.N., was supported by the European Union's Horizon Europe programme under the Marie Skłodowska-Curie Actions (MSCA) grant agreement No. 101154423, project

AIDDRI. P.B., V.B. and K.T. were supported by the project FerrMion of the Ministry of Education, Youth and Sports, Czech Republic, co-funded by the European Union (CZ.02.01.01/00/22_008/0004591).

**Data Availability**

The dataset to reproduce the findings can be accessed through the Zenodo repository link, https://doi.org/10.5281/zenodo.17672235.

**References**

[1]   F. Witte, The history of biodegradable magnesium implants: A review, Acta Biomater. 6 (2010) 1680–1692. https://doi.org/10.1016/J.ACTBIO.2010.02.028.

[2]   Y.F. Zheng, X.N. Gu, F. Witte, Biodegradable metals, Materials Science and Engineering: R: Reports 77 (2014) 1–34. https://doi.org/10.1016/J.MSER.2014.01.001.

[3]   M.P. Staiger, A.M. Pietak, J. Huadmai, G. Dias, Magnesium and its alloys as orthopedic biomaterials: A review, Biomaterials 27 (2006) 1728–1734. https://doi.org/10.1016/j.biomaterials.2005.10.003.

[4]   F. Xing, S. Li, D. Yin, J. Xie, P.M. Rommens, Z. Xiang, M. Liu, U. Ritz, Recent progress in Mg-based alloys as a novel bioabsorbable biomaterials for orthopedic applications, Journal of Magnesium and Alloys 10 (2022) 1428–1456. https://doi.org/10.1016/J.JMA.2022.02.013.

[5]   H. Zhao, J. Cheng, C. Zhao, M. Wen, R. Wang, D. Wu, Z. Wu, F. Yang, L. Sheng, The Recent Developments of Thermomechanical Processing for Biomedical Mg Alloys and Their Clinical Applications, Materials 18 (2025). https://doi.org/10.3390/ma18081718.

[6]   R. Ramprasad, R. Batra, G. Pilania, A. Mannodi-Kanakkithodi, C. Kim, Machine learning in materials informatics: Recent applications and prospects, NPJ Comput. Mater. 3 (2017). https://doi.org/10.1038/s41524-017-0056-5.

[7]   A. Agrawal, A. Choudhary, Deep materials informatics: Applications of deep learning in materials science, MRS Commun. 9 (2019) 779–792. https://doi.org/10.1557/mrc.2019.73.

[8]   Z. Rao, P.-Y. Tung, R. Xie, Y. Wei, H. Zhang, A. Ferrari, T.P.C. Klaver, F. Körmann, P. Thoudden Sukumar, A. Kwiatkowski Da Silva, Y. Chen, Z. Li, D. Ponge, J. Neugebauer, O. Gutfleisch, S. Bauer, D. Raabe, Machine learning-enabled high-entropy alloy discovery, Science (1979). 378 (2022) 78–85. https://www.science.org.

[9]   V. Nandal, S. Dieb, D.S. Bulgarevich, T. Osada, T. Koyama, S. Minamoto, M. Demura, Artificial intelligence inspired design of non-isothermal aging for γ–γ′ two-phase, Ni–Al alloys, Sci. Rep. 13 (2023) 12660. https://doi.org/10.21203/rs.3.rs-2593940/v1.

[10]  V. Nandal, S. Dieb, D.S. Bulgarevich, T. Osada, T. Koyama, S. Minamoto, M. Demura, Analysis of artificial intelligence-discovered patterns and expert-designed aging patterns for 0.2 % proof stress in Ni-Al alloys with γ – γ' two-phase structure, Next Materials 8 (2025) 100564. https://doi.org/10.1016/J.NXMATE.2025.100564.

[11]  H.K.D.H. Bhadeshia, Neural networks and information in materials science, Stat. Anal. Data Min. 1 (2009) 296–305. https://doi.org/10.1002/sam.10018.

[12]  M. Deif, H. Attar, M. Aljaidi, A. Alsarhan, D. Al-Fraihat, A. Solyman, Machine learning alloying design of biodegradable zinc alloy for bone implants using XGBoost and Bayesian optimization, Intelligent Systems with Applications 27 (2025) 200549. https://doi.org/10.1016/J.ISWA.2025.200549.

[13]  K. Aas, M. Jullum, A. Løland, Explaining individual predictions when features are dependent: More accurate approximations to Shapley values, Artif. Intell. 298 (2021) 103502. https://doi.org/10.1016/J.ARTINT.2021.103502.

[14]  C. Zhang, Y. Zhang, B. Ren, Y. Wu, Y. Hu, Y. Chai, L. Xu, Q. Wang, Accelerated design of age-hardened Mg-Ca-Zn alloys with enhanced mechanical properties via machine learning, Comput. Mater. Sci. 249 (2025) 113665. https://doi.org/10.1016/J.COMMATSCI.2025.113665.

[15]  M.K. Guru, J. Bohlen, R.C. Aydin, N. Ben Khalifa, Machine learning pipeline for Structure–Property modeling in Mg-alloys using microstructure and texture descriptors, Acta Mater. 295 (2025) 121132. https://doi.org/10.1016/J.ACTAMAT.2025.121132.

[16]  S. min Ai, D. ran Fang, Y. wei Guo, J. Ye, X. ping Lin, Machine learning-driven prediction of microstructure-mechanical property relationships in Mg-Al alloys, J. Alloys Compd. 1036 (2025) 181995. https://doi.org/10.1016/J.JALLCOM.2025.181995.

[17]  L. Breiman, Random Forests, Mach. Learn. 45 (2001) 5–32.

[18]  T. Chen, C. Guestrin, XGBoost: A scalable tree boosting system, in: Proceedings of the ACM SIGKDD International Conference on Knowledge Discovery and Data Mining, Association for Computing Machinery, 2016: pp. 785–794. https://doi.org/10.1145/2939672.2939785.

[19]  L. Prokhorenkova, G. Gusev, A. Vorobev, A.V. Dorogush, A. Gulin, CatBoost: unbiased boosting with categorical features, (2019). http://arxiv.org/abs/1706.09516.

[20]  Y. Ding, C. Wen, P. Hodgson, Y. Li, Effects of alloying elements on the corrosion behavior and biocompatibility of biodegradable magnesium alloys: A review, J. Mater. Chem. B 2 (2014) 1912–1933. https://doi.org/10.1039/c3tb21746a.

[21]  S. Wang, H. Pan, D. Xie, D. Zhang, J. Li, H. Xie, Y. Ren, G. Qin, Grain refinement and strength enhancement in Mg wrought alloys: A review, Journal of Magnesium and Alloys 11 (2023) 4128–4145. https://doi.org/10.1016/J.JMA.2023.11.002.

[22] C.H. Cáceres, D.M. Rovera, Solid solution strengthening in concentrated Mg–Al alloys, Journal of Light Metals 1 (2001) 151–156. https://doi.org/10.1016/S1471-5317(01)00008-6.

[23] X. Wu, X. Jing, H. Xiao, S. Ouyang, A. Tang, P. Peng, B. Feng, M. Rashad, J. She, X. Chen, K. Zheng, F. Pan, Controlling grain size and texture in Mg–Zn–Mn alloys from the interaction of recrystallization and precipitation, Journal of Materials Research and Technology 21 (2022) 1395–1407. https://doi.org/10.1016/j.jmrt.2022.09.108.

[24] H. Wang, X.C. Luo, D.T. Zhang, C. Qiu, D.L. Chen, High-strength extruded magnesium alloys: A critical review, J. Mater. Sci. Technol. 199 (2024) 27–52. https://doi.org/10.1016/J.JMST.2024.01.089.

[25] J. Singh, A. Wahab Hashmi, S. Ahmad, Y. Tian, Critical review on biodegradable and biocompatibility magnesium alloys: Progress and prospects in bio-implant applications, Inorg. Chem. Commun. 169 (2024) 113111. https://doi.org/10.1016/J.INOCHE.2024.113111.

[26] J. She, P. Peng, L. Xiao, A.T. Tang, Y. Wang, F.S. Pan, Development of high strength and ductility in Mg–2Zn extruded alloy by high content Mn-alloying, Materials Science and Engineering: A 765 (2019). https://doi.org/10.1016/j.msea.2019.138203.

[27] B. sheng LIU, M. miao CAO, Y. zhong ZHANG, Y. HU, C. wei GONG, L. feng HOU, Y. hui WEI, Microstructure, anticorrosion, biocompatibility and antibacterial activities of extruded Mg−Zn−Mn strengthened with Ca, Transactions of Nonferrous Metals Society of China (English Edition) 31 (2021) 358–370. https://doi.org/10.1016/S1003-6326(21)65501-2.

[28] E. Zhang, D. Yin, L. Xu, L. Yang, K. Yang, Microstructure, mechanical and corrosion properties and biocompatibility of Mg-Zn-Mn alloys for biomedical application, Materials Science and Engineering C 29 (2009) 987–993. https://doi.org/10.1016/j.msec.2008.08.024.

[29] A.R. Khan, N.S. Grewal, C. Zhou, K. Yuan, H.J. Zhang, Z. Jun, Recent advances in biodegradable metals for implant applications: Exploring in vivo and in vitro responses, Results in Engineering 20 (2023) 101526. https://doi.org/10.1016/J.RINENG.2023.101526.

[30] K. Tesař, J. Luňáčková, M. Jex, M. Žaloudková, R. Vrbová, M. Bartoš, P. Klein, L. Vištejnová, J. Dušková, E. Filová, Z. Sucharda, M. Steinerová, S. Habr, K. Balík, A. Singh, In vivo and in vitro study of resorbable magnesium wires for medical implants: Mg purity, surface quality, Zn alloying and polymer coating, Journal of Magnesium and Alloys 12 (2024) 2472–2488. https://doi.org/10.1016/J.JMA.2024.06.003.